\documentclass[twocolumn,aps,prd,showpacs,nofootinbib,superscriptaddress,
preprintnumbers,floatfix,10pt]{revtex4-1}

\usepackage{amsmath}
\usepackage{amsfonts}
\usepackage{amssymb}
\usepackage{epsfig}
\usepackage{graphicx}
\usepackage{color}
\usepackage{slashed}
\usepackage{epstopdf}
\usepackage{dsfont}
\usepackage{bbold}

\usepackage{hyperref}

\newcommand{\tr}{\text{tr}}
\newcommand{\STr}{\text{STr}}
\newcommand{\Tr}{\text{Tr}}
\newcommand{\Eqref}[1]{Eq.~\eqref{#1}}
\newcommand{\Nc}{N_{\mathrm{c}}}
\newcommand{\Nf}{N_{\mathrm{f}}}
\newcommand{\pat}{\partial_t}

\DeclareMathOperator{\Obig}{O}

\newcommand{\D}{D} 
\DeclareMathOperator{\SU}{SU}

\begin{document}

\preprint{}

\title{Quantum field theories of relativistic Luttinger fermions} 

\author{Holger Gies}
\email{holger.gies@uni-jena.de}
\affiliation{Theoretisch-Physikalisches Institut, 
Abbe Center of Photonics, Friedrich Schiller University Jena, Max Wien 
Platz 1, 07743 Jena, Germany}
\affiliation{Helmholtz-Institut Jena, Fr\"obelstieg 3, D-07743 Jena, Germany}
\affiliation{GSI Helmholtzzentrum für Schwerionenforschung, Planckstr. 1, 
64291 Darmstadt, Germany} 
\author{Philip Heinzel}
\email{philip.heinzel@uni-jena.de}
\affiliation{Theoretisch-Physikalisches Institut, 
Abbe Center of Photonics, Friedrich Schiller University Jena, Max Wien 
Platz 1, 07743 Jena, Germany}
\author{Johannes Laufkötter}
\email{johannes@laufkoetter-online.de}
\affiliation{Theoretisch-Physikalisches Institut, 
Abbe Center of Photonics, Friedrich Schiller University Jena, Max Wien 
Platz 1, 07743 Jena, Germany}
\author{Marta Picciau}
\email{marta.picciau@uni-jena.de}
\affiliation{Theoretisch-Physikalisches Institut, 
Abbe Center of Photonics, Friedrich Schiller University Jena, Max Wien 
Platz 1, 07743 Jena, Germany}

\begin{abstract}
We propose relativistic Luttinger fermions as a new ingredient for the 
construction of fundamental quantum field theories. We construct the 
corresponding Clifford algebra and the spin metric for relativistic invariance 
of the action using the spin-base invariant formalism. The corresponding 
minimal spinor has 32 complex components, matching with the degrees of freedom 
of a standard-model generation including a right-handed neutrino. 
The resulting fermion fields exhibit a canonical scaling different 
from Dirac fermions and thus support the construction of novel relativistic and 
perturbatively renormalizable, interacting quantum field theories. In 
particular, new asymptotically free self-interacting field theories can be 
constructed, representing first examples of high-energy complete quantum field 
theories based on pure matter degrees of freedom. Gauge theories with 
relativistic Luttinger fermions exhibit a strong paramagnetic dominance, 
requiring large nonabelian gauge groups to maintain asymptotic freedom. 
We comment on the possibility to use Luttinger fermions for particle physics 
model building and the expected naturalness properties of such models.

\end{abstract}


\maketitle


\section{Introduction}
\label{intro}

One of the remarkable features of quantum field theories is given by the 
interconnection of fields as representations of the Lorentz group 
\cite{Wigner:1939cj,Bargmann:1948ck}, their powercounting dimensionality and 
the renormalizability of interacting field theories in $d=3+1$-dimensional 
spacetime \cite{Wilson:1971dh,ZinnJustin:1989mi,Weinberg:1995mt}. These 
interconnections are particularly obvious in the standard model containing spin 
$0,\frac{1}{2},1$ fields and accommodating all possible renormalizable 
interactions allowed by the symmetries. 

In addition to the experimental searches for further degrees of freedom, 
theoretical studies of consistent quantum field theories have paved the way for 
new concepts, e.g., with a consistent quantization of interacting spin 
$\frac{3}{2}$ particles requiring supersymmetry. 

In the present paper, we explore the possibility to construct relativistic 
versions of Luttinger fermions and perform first perturbative studies of
corresponding interacting quantum field theories. This type of fermions has 
been discovered by Luttinger while searching for the most general form of a 
nonrelativistic Hamiltonian of a semiconductor excitation in a magnetic 
field \cite{LuttingerPhysRev.102.1030}. In recent solid-state research, 
these nonrelativistic degrees of freedom find extensive application in 
spin-orbit coupled materials with quadratic band touching/crossing points 
(e.g., inverted band gap semiconductors, pyrochlore iridates) 
\cite{Murakami:2004zz,Moon:2012rx,Savary:2014gka}; such systems can give rise 
to interesting quantum critical phenomena 
\cite{Herbut:2014lfa,Janssen:2015xga,Janssen:2015jga,Boettcher:2016wft,
Janssen:2016xvc,Boettcher:2016iiv, Ray:2018gtp,Ray:2020mlg,Ray:2021moi, 
Ray:PhD}. Also gauged versions have been studied recently in the context of 
quantum spin liquids \cite{Dey:2022lkx}.

While the generalization of the underlying algebra to the relativistic case is,
in principle, straightforward, we find that the construction of a fully
relativistic action requires a reducible representation in terms of the related
Dirac algebra with interesting consequences for the construction of interacting
quantum field theories. From the viewpoint of the propagator pole structure, the
theories exhibit typical features of a higher-derivative theory
\cite{Pais:1950za,Lee:1970iw,Stelle:1976gc,Grinstein:2007mp}, whereas the
ultraviolet (UV) properties of loop integrals resembles that of standard scalar
field theories with the decisive difference that self-interacting theories can
be asymptotically free. 

\section{Relativistic Luttinger fermions}

We propose the kinetic action for a relativistic theory  with 
Luttinger fermions in $d$ dimensional spacetime to be of the form
\begin{equation}
 S=\int d^d x \left[ \bar\psi G_{\mu\nu} (i\partial^\mu) (i \partial^\nu) \psi 
\right],
\label{eq:Skin}
\end{equation}
where $\psi$ denotes the Grassmann-valued spinor, $\bar\psi$ its conjugate to be constructed, 
and $G_{\mu\nu}$ is a $d_\gamma\times d_\gamma$ dimensional matrix for each 
fixed pair of Lorentz indices $\mu,\nu=0, \dots, (d-1)$. In order to remove the 
Lorentz-reducible part proportional to a trivial Klein-Gordon operator, the 
Luttinger matrices $G_{\mu\nu}$ are Lorentz traceless, $G^\mu{}_\mu=g^{\mu\nu} 
G_{\mu\nu}=0$. Also, they are Lorentz symmetric $G_{\mu\nu}=G_{\nu\mu}$ as 
is obvious from \Eqref{eq:Skin}, satisfying the anti-commuting algebra
\begin{equation}
 \{G_{\mu\nu}, G_{\kappa\lambda}\} = - \frac{2}{d-1} 
g_{\mu\nu}g_{\kappa\lambda}+\frac{d}{d-1} (g_{\mu\kappa} g_{\nu\lambda} + 
g_{\mu\lambda}g_{\nu\kappa}),
\label{eq:Galgebra}
\end{equation} 
generalizing the Abrikosov algebra for the spatial Euclidean   
\cite{Abrikosov:1974a,Janssen:2015xga} to the Lorentzian case.
The tensor structure of the right-hand side is fixed by Lorentz and 
index symmetries. The prefactors follow from the Lorentz tracelessness of
$G_{\mu\nu}$ and from the requirement that the Luttinger operator should 
square to the square of the Klein-Gordon operator,
\begin{equation}
 G_{\mu\nu} (i\partial^\mu) (i \partial^\nu) G_{\kappa\lambda} 
(i\partial^\kappa) (i \partial^\lambda) = (\partial^2)^2.
\label{eq:Gsquared}
\end{equation}
With $G_{\mu\nu}$ being symmetric in $\mu,\nu$ and traceless, we need at least 
$d_{\text{e}}= \frac{1}{2} d(d+1)-1$ linearly independent elements to span the 
algebra \eqref{eq:Galgebra}. Since \Eqref{eq:Galgebra} defines a Clifford 
algebra, the dimension $d_\gamma$ must at least be that of the irreducible 
representation $d_{\gamma,\text{irr}}=2^{\lfloor d_{\text{e}}/2\rfloor}$. 
Naively, this suggests that we need a $d_{\gamma,\text{irr}}=16$ dimensional 
representation for the required $d_{\text{e}}=9$ elements $G_{\mu\nu}$ in 
$d=4$ spacetime dimensions. Using a metric $g=\text{diag}(+,-,-,\dots)$, the 
$G_{0i}$ are anti-hermitean with respect to their spin indices while all 
other $G_{ik}$ and $G_{00}$ are hermitean. In $d=4$ spacetime dimensions, we can 
use a corresponding $d_\gamma$ dimensional Euclidean Dirac algebra 
$\{\gamma_A,\gamma_B\}=2\delta_{AB}$ with $A,B\in 1, \dots, 9$ to construct an 
explicit representation of the $G_{\mu\nu}$ as linear combinations of the 
$\gamma_A$ (see App. \ref{sec:AppA} for an in-depth discussion of the relativistic Abrikosov algebra and its representation).  So far, the 
construction is analogous to Luttinger 
fermion applications in condensed matter physics, replacing the spatial 
Euclidean metric by the Minkowski metric, cf. \cite{Janssen:2015xga}. 

For the relativistic action, we also need the definition of the conjugate 
spinor $\bar\psi$. A unitary time evolution requires a real action which 
suggests to write
$\bar\psi=\psi^\dagger h$, where $h$ denotes the spin metric. For its 
construction, we note that the Abrikosov algebra \eqref{eq:Galgebra} is 
invariant under 
Lorentz transformations with respect to the Lorentz indices as well as 
invariant under SL($d_\gamma,\mathbb{C}$) spin-base transformations 
\cite{Schroedinger:1932a,Bargmann:1932a,Weldon:2000fr,Gies:2013noa} 
\begin{equation}
 G_{\mu\nu}\to \mathcal{S} G_{\mu\nu} \mathcal{S}^{-1}, \quad \mathcal{S}\in 
\text{SL}(d_\gamma,\mathbb{C}).
\label{eq:spinbase}
\end{equation}
The action is spin-base invariant with $\psi\to \mathcal{S} \psi$, provided 
that the spin metric transforms as $h\to (\mathcal{S}^\dagger)^{-1} h 
\mathcal{S}^{-1}$. The condition that $\bar\psi \psi$ should form a real scalar 
(mass term) implies $h^\dagger=h$. The kinetic term \eqref{eq:Skin} is real if
\begin{equation}
 \{h,G_{0i}\}=0, \quad [h,G_{ij}]=0, \quad 
[h,G_{\underline{\mu}\underline{\mu}}]=0,
\label{eq:hcomms}
\end{equation}
where underscored indices are not summed over. Now, the important point is that 
no solution for $h$, satisfying \Eqref{eq:hcomms} exists in the irreducible 
representation $d_{\gamma,\text{irr}}=16$ for $d=4$. The construction of a 
relativistic theory of Luttinger fermions thus requires the use of a
reducible representation of the corresponding Euclidean Dirac Clifford algebra, 
e.g., $d_\gamma=32$ as the minimal possibility in $d=4$ dimensional spacetime. 
For instance, for a representation with $G_{0i}= i \sqrt{\frac{2}{3}} 
\gamma_{A=i}$, a suitable spin metric is given by 
$h=\gamma_1\gamma_2\gamma_3\gamma_{10}$. (Another linearly independent solution 
is given by replacing $\gamma_{10}$ by $\gamma_{11}$.)

In summary, the action \eqref{eq:Skin} defines a free theory of 
massless propagating relativistic Luttinger fermions. These fermions have 
$d_\gamma=32$ complex components that obey the classical equation of motion 
$G_{\mu\nu}\partial^\mu\partial^\nu \psi=0$. Because of \Eqref{eq:Gsquared}, 
each component satisfies the Klein-Gordon equation. The theory is 
SL($d_\gamma,\mathbb{C}$) spin-base invariant.

\section{Self-interacting quantum Luttinger fields}
Because of the kinetic term being quadratic in the derivatives, the canonical
mass dimension of the fermions is $[\psi]=1$ in $d=4$ analogous to a scalar
field. Therefore, quartic self-interactions are perturbatively renormalizable.
With 1024 independent bilinears (compared to 16 for Dirac fermions), there are
plenty of opportunities for model building. Here, we concentrate on a few
selected channels in the Euclidean domain, starting with the Luttinger variant of the Gross-Neveu model
\cite{Gross:1974jv} in $d=4$. In contradistinction to Dirac fermions, the
Gross-Neveu channel generates also a tensor channel such that a Fierz-complete
basis of local interactions is given by
\begin{equation}
 S=\int d^4x \left[-\bar\psi G_{\mu\nu}\partial^\mu \partial^\nu \psi + 
\frac{\lambda_0}{2} (\bar\psi\psi)^2 + \frac{\lambda_t}{2} 
(\bar\psi G_{\mu\nu}\psi)^2\right]. 
\label{eq:GrossNeveu}
\end{equation}
As a first step towards the quantum analysis of this field theory, we compute 
the one-loop $\beta$ functions, yielding 
\begin{eqnarray}
 \partial_t \lambda_0 &=& -\frac{1}{(12\pi)^2}\big( 540 \lambda_0^2- 
528\lambda_0\lambda_t\big)\label{eq:patlambda0},\\
 \partial_t \lambda_t &=& -\frac{1}{(12\pi)^2}\big( - 992\lambda_t^2 + 28 
\lambda_0\lambda_t -3 \lambda_0^2 \big).\label{eq:patlambdat}
\end{eqnarray}
Details of the calculation are provided in App.~\ref{sec:AppB}.
For the pure Gross-Neveu case, $\lambda_t=0$, the scalar coupling 
$\lambda_0$ has a negative $\beta$ function \eqref{eq:patlambda0} exhibiting 
asymptotic freedom. This is in agreement with naive expectations, as the standard 
Gross-Neveu model is asymptotically free in its critical dimension. For Dirac 
fermions, the critical dimension is $d_{\text{cr}}=2$; but for Luttinger 
fermions, we have $d_{\text{cr}}=4$. In comparison to a scalar $\phi^4$ 
theory which is not asymptotically free, we deal here with diagrams of the same 
topology, but the fermion loop comes with another minus sign.

\begin{figure}[t]
\includegraphics[width=0.48\textwidth]{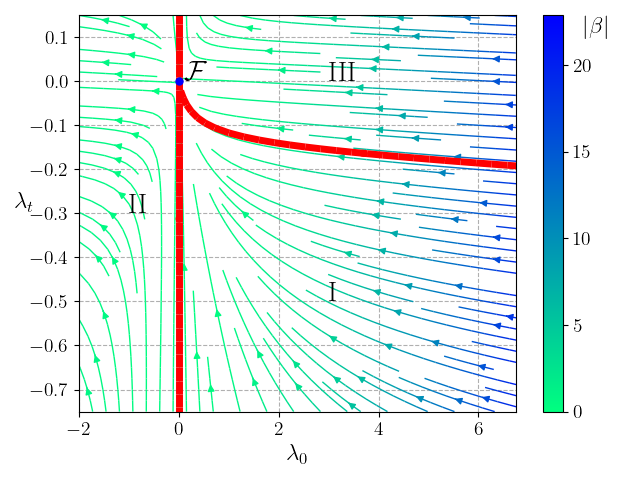}
\caption{Phase diagram of the Gross-Neveu-like model with 
relativistic Luttinger fermions in $d=4$ for $\Nf=1$ and $d_\gamma=32$ with 
arrows indicating the flow towards the UV. In region I, The model is 
asymptotically free in both the scalar coupling $\lambda_0$ as well as the 
tensor channel $\lambda_t$, approaching the Gaussian fixed point $\mathcal{F}$ 
logarithmically.}
\label{fig:lambdaplane}
\end{figure}

However, the tensor channel is generated by the scalar channel, cf. 
\Eqref{eq:patlambdat}. Therefore, the analysis should be performed in the 
$(\lambda_0,\lambda_t)$ plane. The phase diagram reveals that the Gaussian 
fixed point is UV attractive in a large region with $\lambda_0>0$ and 
$\lambda_t$ sufficiently 
negative, see Fig.~\ref{fig:lambdaplane}. Defining the angle $\alpha =\arctan 
\frac{\lambda_t}{\lambda_0}$, the model is asymptotically free for $-90^\circ 
\leq \alpha \lesssim 0^\circ$ with the scalar coupling dominating for $\alpha> 
-45.8^\circ$. To our knowledge, this is the first example of an asymptotically 
free, UV complete, pure fermionic matter theory.

As a second example, we consider a fermionic model with a continuous 
chiral/axial symmetry. Analogously to the Dirac case, we need an algebra 
element that anti-commutes with all $G_{\mu\nu}$ as well as with $h$. For the 
present case, this element is given by $\gamma_{10}$, and it is straightforward 
to check that $\psi\to e^{i\vartheta\gamma_{10}}\psi$, $\bar\psi\to\bar\psi 
e^{i\vartheta\gamma_{10}}$ is a U(1) ``axial'' symmetry of the kinetic action. 
A self-interacting model with this symmetry is given by the analogue of the 
Nambu--Jona-Lasinio (NJL) model \cite{Nambu:1961tp},
\begin{equation}
 S=\int d^4x \left[-\bar\psi G_{\mu\nu}\partial^\mu \partial^\nu \psi + 
\frac{\lambda}{2}[(\bar\psi\psi)^2-(\bar\psi \gamma_{10}\psi)^2]\right].
\end{equation}
The one-loop $\beta$ function for the NJL coupling yields for $d_\gamma=32$, cf. App. \ref{sec:AppB},
\begin{equation}
 \partial_t \lambda =  -\frac{4\Nf}{\pi^2} \lambda^2 \label{eq:patlambda}.
\end{equation}
Again, we observe that the coupling is asymptotically free. Here, we have 
neglected further vector/axial-vector channels that are generated by the NJL 
coupling. This is justified in the limit of large fermion flavor number $\Nf$ 
where the channels decouple. 

In summary, self-interacting quantum field theories with Luttinger fermions 
give rise to asymptotically free RG trajectories emanating from the Gaussian 
fixed point thus representing UV complete theories. Diagrammatically, this is 
similar to Dirac fermionic models in their critical dimension. For Luttinger 
fermions, we have $d_{\text{cr}}=4$ guaranteeing perturbative 
renormalizability. Conversely, the theories become strongly interacting towards 
the infrared (IR) such that phenomena such as dynamical symmetry breaking and 
mass generation can be expected. The exploration of the long-range phase 
structure of the present and similar models will be an interesting field of 
future research.

\section{Connection to Dirac fermions}

Counting the degrees of freedom, a 32-component Luttinger spinor contains eight 
4-component Dirac spinors. Noteworthily, this covers the number of spinor 
degrees of freedom in one standard-model family: up and down quarks with 3 
colors each, an electron and a neutrino (including a possible right-handed 
component). This raises the question as to whether a UV-completion of the 
standard model in terms of a model with Luttinger-fermionic matter can be 
constructed. 

For this, a minimum requirement is that the spin-base symmetry 
SL$(32,\mathbb{C})$ needs to be broken down to the Dirac spin-base symmetry 
SL$(4,\mathbb{C})$ (possibly times some residual Dirac flavor symmetry). The 
Gross-Neveu- or NJL-type models presumably preserve spin-base symmetry also 
across possible strong-coupling transitions. Therefore, an explicit symmetry 
breaking mechanism appears more attractive -- also in order to avoid a 
potentially large number of Goldstone bosons. 

Explicit breaking terms could be formulated on the level of RG marginal
four-fermion interactions. However, such marginal terms then induce also an RG
relevant term which is given by the Dirac kinetic term $\sim \zeta_{\text{D}}
\int_x \bar\psi  P_{\text{D}}(i\slashed{\partial}) \psi$, where
$P_{\text{D}}(i\slashed{\partial})$ is a suitable Abrikosov algebra element
linear in the Dirac operator $i\slashed{\partial}$ projecting the Luttinger
components onto Dirac components. The prefactor $\zeta_{\text{D}}$ is a coupling
of mass dimension one. Powercounting suggests that the UV remains still
dominated by the RG behavior of relativistic Luttinger theory, whereas the
fermion propagators become Dirac-like at momenta below $\zeta_{\text{D}}$. Such
a transition from Luttinger to Dirac fermions has been studied in the
nonrelativistic case in \cite{Ray:2018gtp} for Bernal-stacked bilayer honeycomb
lattices such as bilayer graphene; for a scalar analogue, see
\cite{Buccio:2022egr}.

The explicit Dirac term also comes with another advantage: While we have 
studied theories of massless Luttinger fermions so far, a massive 
generalization of the free field equation reads
\begin{equation}
 (-G_{\mu\nu}\partial^\mu\partial^\nu-m^2)\psi=0. 
 \label{eq:massieeq}
\end{equation}
Transition to momentum space and using \Eqref{eq:Gsquared} yields
$0=(p^4-m^4)\psi(p) = (p^2-m^2)(p^2+m^2)\psi(p)$. This illustrates that each
component of the Luttinger spinor satisfies the classical field equation of a
higher-derivative theory
\cite{Pais:1950za,Lee:1970iw,Stelle:1976gc,Grinstein:2007mp,Woodard:2015zca}.
This observation also suggests that the massive theory features tachyonic
solutions with a negative spectral weight in addition to conventional massive
modes. These are consequences of Ostrogradsky's theorem implying that
Hamiltonians of higher-derivative theories are unbounded from below
\cite{Ostrogradsky:1850fid}. Whereas this appears to point to instabilities (or
nonunitarity) at the quantum level, such theories are nevertheless discussed
intensely in the literature in a variety of contexts both on the classical and
the quantum level. Concrete proposals for a consistent treatment on the quantum
level have been suggested
\cite{Lee:1970iw,Narnhofer:1978sw,Hawking:2001yt,Bender:2007wu,Grinstein:2007mp,Garriga:2012pk,Salvio:2014soa,Smilga:2017arl,Becker:2017tcx,Anselmi:2018kgz,Gross:2020tph,Donoghue:2021eto,Platania:2019qvo}.
Interestingly, explicit proofs exist for classical example systems that their
motion remains stable for all initial conditions, despite interactions with the
seemingly unstable modes \cite{Deffayet:2021nnt,Deffayet:2023wdg}. 

Even if instabilities from negative-energy modes persist for the present
class of models in a detailed quantum analysis, they can still be discussed from
the perspective of effective field theory (EFT), as long as the rate of
instability is small enough to satisfy all relevant phenomenological bounds. For
instance, a fairly universal bound arises from gravity-mediated vacuum decay
(into photons and negative-energy modes) which would contribute to the
astrophysical diffuse photon background \cite{Cline:2003gs,Cline:2023hfw}. In
fact, cosmology and its puzzles involving dark energy and dark matter have been
a fruitful field for the application of field theories with negative energy
modes \cite{Caldwell:1999ew}. A quantum treatment of such theories in an EFT
framework bears the possibility of ameliorating tensions in current cosmological
data \cite{Cline:2023cwm,Cline:2024zhs}, or provides mechanisms to produce
dark-matter candidates accessible to direct detection
\cite{Cline:2023hfw,Cline:2024zhs}. Whether or not theories with Luttinger
fermions can be useful in this context is a subject for future research.

For the purpose of this work,
we emphasize that we have paid careful attention to possible problems arising
from the in-principle unboundedness of the Hamiltonian, see Appendices \ref{sec:AppB} and \ref{sec:AppC}, but found no
impact on the quantities under study. In this context, we note that an explicit
Dirac kinetic term has the potential to decouple the tachyonic poles from
the real momentum axis, as the Dirac equation admits only classical solutions
with $p^2=m^2$. We thus consider an explicit Dirac kinetic term as a useful
ingredient for future model building as well as for an investigation of
the implications of Ostrogradsky's theorem.

\section{Gauge theories with Luttinger fermions}
\label{sec:gauge}

Luttinger fermions can straightforwardly be coupled to gauge fields by 
minimal coupling. For instance, the action for quantum electrodynamics (QED) 
with Luttinger fermions reads
\begin{equation}
 S=\int_x \left[ -\frac{1}{4} F_{\mu\nu}F^{\mu\nu} - \bar\psi G_{\mu\nu}D^\mu 
D^\nu\psi -m^2 \bar\psi\psi\right],
\label{eq:SLQED}
\end{equation}
where $D^\mu=\partial^\mu - i e A^\mu$. We have included here also a mass term. 
As we are ultimately interested in the one-loop $\beta$ function in a 
mass-independent scheme, the tachyonic modes mentioned in the previous 
section are of no relevance for the present study. 

A convenient way to compute the one-loop $\beta$ function of the running gauge 
coupling proceeds via the one-loop effective action upon integrating out the 
fermions:
\begin{eqnarray}
\Gamma_{1\ell}[A]&=& -i \ln \det (-G_{\mu\nu}D^\mu D^\nu -m^2) \nonumber\\
&=& 
-\frac{i}{2} \ln\det [-(G_{\mu\nu}D^\mu D^\nu)^2 + m^4)],
\label{eq:det}
\end{eqnarray}
where in the last step we have used $\gamma_{10}$-hermiticity of the kinetic
term, $\gamma_{10} G_{\mu\nu}D^\mu D^\nu \gamma_{10}= -G_{\mu\nu}D^\mu D^\nu$.
An interesting and relevant structure arising from the minimally coupled kinetic
term is given by the spin-field coupling, reading $\sim \frac{ie}{2}
[G_{\mu\nu},G_{\kappa\lambda}] F^{\nu\lambda} \{D^\mu,D^\kappa\} +
\mathcal{O}(\partial F)$, see App.~\ref{sec:AppC} for details. This is the
analogue of the Pauli term $\sim - \frac{e}{2} \sigma_{\mu\nu} F^{\mu\nu}$ for
Dirac spinors, leading to an enhancement of ``paramagnetic'' contributions. 

Expanding the determinant in powers of the field strength, the leading order
term $\sim F_{\mu\nu} F^{\mu\nu}$ contains the information about the
renormalization of the photon wave function; this computation is presented in
App.~\ref{sec:AppC}. Within the background field formalism, the wave-function
renormalization is connected to the renormalization of the coupling
\cite{Dittrich:1985yb,Gies:2002af}, yielding the $\beta$ function
\begin{equation}
 \pat e^2 = \frac{4\cdot 19}{9 \pi^2} \, e^4 = \frac{4}{9\pi^2} \left( 22 
\Big|_{\text{para}}- 3\Big|_{\text{dia}}\right) e^4.
\label{eq:betaQED}
\end{equation}
In the last expression, we have decomposed the result into a 
``diamagnetic'' contribution from the Klein-Gordon operator contained in 
\Eqref{eq:det} and a ``paramagnetic'' contribution arising from the remaining 
terms including the spin-field coupling. Obviously, we observe 
\textit{paramagnetic dominance}, i.e. the paramagnetic 
contributions dominate the final result and are also responsible for the sign 
of the $\beta$ function, as is known for many theories including nonabelian 
gauge theories and even gravity \cite{Nink:2012vd}. 

Due to this strong paramagnetic dominance, the $\beta$ function is positive. As 
in ordinary QED, the theory is thus not asymptotically free, and the 
vicinity of the Gaussian fixed point does not support RG trajectories that are 
UV complete and yield an interacting theory in the long-range limit. As for the 
standard model, a different mechanism is needed to establish high-energy 
completeness. 

Let us finally turn to nonabelian gauge theories, generalizing the action 
\eqref{eq:SLQED} to a nonabelian SU($\Nc$) gauge group with $\Nc$ colors. Since 
the Yang-Mills sector remains unmodified, the resulting one-loop $\beta$ 
function for the coupling $g$ of QCD with $\Nf$ relativistic Luttinger quarks 
can immediately be written down,
\begin{equation}
 \pat g^2 = - \frac{1}{3\pi^2} \left( \frac{11}{8}\Nc - \frac{2\cdot 19}{3} \Nf 
\right) g^4.
\label{eq:betaLQCD}
\end{equation}
As is the standard case, asymptotic freedom of QCD can get lost for a large 
number of fermionic degrees of freedom relative to the number of colors. Due to 
the strong paramagnetic dominance of the Luttinger fermions, the 
critical color number below which asymptotic freedom is lost is comparatively 
large:
\begin{equation}
 \Nc{}_{\text{,cr}}= \frac{304}{33} \Nf \simeq 9.21 \Nf.
 \label{eq:Nccr}
\end{equation}
Hence, for $\Nf=1$ Luttinger flavors, QCD is asymptotically free for 
SU($\Nc\geq 10$).

An inclusion of all fermionic matter of the standard model (including
right-handed neutrinos) would require $\Nf=3$, since a standard-model generation
fits into a single Luttinger flavor, i.e., the number of generations equals the
number of flavors. Correspondingly, asymptotically free grand unified models
based on a simple Lie group can be constructed for all SU($\Nc\geq 28$),
provided all obstructions can be met
\cite{Maas:2015gma,Maas:2016ngo,Maas:2017xzh,Sondenheimer:2019idq}. In
principle, it is conceivable that this gauge group can be broken dynamically by
suitable fermionic condensates that are seeded by four-fermion interactions
instead of explicit independent Higgs fields. While this would be similar in
spirit to models of top-quark condensation
\cite{Miransky:1988xi,Miransky:1989ds,Bardeen:1989ds,Hill:2002ap}, the essential
difference is that the four fermion interactions are RG marginal and
perturbatively renormalizable -- and potentially asymptotically free.

\section{Conclusions}
\label{sec:conc}

We have constructed relativistic versions of Luttinger fermions in analogy to 
effective low-energy degrees of freedom of nonrelativistic solid-state 
systems. We propose to use these relativistic versions as fundamental degrees 
of freedom of interacting quantum field theories which can straightforwardly be 
constructed using perturbative quantization. 
Owing to their powercounting properties, models with quartic self-interactions 
of relativistic Luttinger fermions can be asymptotically free in $d=4$ 
dimensional spacetime and thus have a chance to exist as fundamental quantum 
field theories on all scales. We provided perturbative evidence for this for 
the analogues of Gross-Neveu and NJL models. Since $d=4$ is the RG 
critical dimension for our new models, the perturbative RG running of the 
couplings is logarithmic and a correspondingly large degree of UV insensitivity 
is obtained. Conversely, these models become strongly interacting 
in the long-range limit and their phase structure presumably 
characterized by dimensional transmutation and condensate formation deserves to 
be explored in detail. 

A crucial ingredient for the construction of these relativistic models is given 
by the spin metric which requires a $d_\gamma=32$ dimensional representation of 
the Abrikosov algebra. It is fascinating to see that one flavor of Luttinger 
fermions can thus host a whole standard-model generation. A corresponding 
perturbative study of gauge theories with Luttinger fermion matter reveals that 
comparatively large gauge groups are needed in order to preserve asymptotic 
freedom.  

A suitable combination of gauge and fermionic self-interactions holds the 
promise to remain asymptotically free and high-energy complete while entailing 
condensate formation and thus dynamical (gauge) symmetry breaking at 
low-energies. Such theories would be \textit{technically natural} 
\cite{tHooft:1979rat,Giudice:2008bi,Grinbaum:2009sk,Dine:2015xga,%
Hossenfelder:2018ikr} and thus of great interest to model building. In 
addition, however, the transition to low energies requires a breaking of the 
Luttinger spinors and the corresponding spin-base symmetry down to Dirac 
spinors. For this, we suggest an explicit breaking through a Dirac kinetic term 
which is RG relevant and comes with a dimensionful coupling. Though the scale 
setting of the latter is not technically natural, the associated sensitivity to 
a high-energy scale is only powercounting linear as opposed to quadratic in the 
standard model.

For a minimal technically natural extension of the standard model, it appears
worthwhile to aim at the construction of a separate Luttinger fermion sector
designed such that it forms a bilinear fermion condensate at low energies with
the quantum numbers of the standard model Higgs field. In this case, the
electroweak scale would be set by the scale of dimensional transmutation of the
Luttinger fermion self-interaction which features the desired logarithmic
sensitivity to the UV physics. Because of the strong paramagnetic dominance such
a model would inevitably exert a strong influence on the (still logarithmic)
running of the electroweak gauge sector possibly at the expense of no asymptotic
freedom. A conclusive analysis requires to study the RG interplay of the gauge
sector with the renormalizable fermionic self-interactions, cf.
\cite{Gies:2003dp}.

To conclude, the new kind of relativistic fermion degrees of freedom pave the 
way to unprecedented explorations of new particle physics models in four 
spacetime dimensions.

\section*{Acknowledgments}

We thank Stefan Flörchinger, Lukas Janssen, Roberto Percacci, Shouryya 
Ray, Laura Schiffhorst, Gian Paolo 
Vacca, Christof Wetterich for valuable discussions.  
This work has been funded by the Deutsche 
Forschungsgemeinschaft (DFG) under Grant No. 406116891 
within the Research Training Group RTG 2522/1 and under Grant Nos.
392856280, 416607684, and 416611371 within the  Research Unit FOR2783/2. 
PH acknowledges funding by the TOQI project of the profile area LIGHT.

\appendix

\section{Relativistic Abrikosov algebra}
\label{sec:AppA}

For the construction of the relativistic kinetic action in \Eqref{eq:Skin}, we 
use the anti-commuting Clifford algebra of elements $G_{\mu\nu}$ acting on the 
Luttinger fermions as it has first been written down for the nonrelativistic 
case by Abrikosov \cite{Abrikosov:1974a}. For 
$\{G_{\mu\nu},G_{\kappa\lambda}\}\sim \mathds{1}$ to transform as a tensor 
under Lorentz transformations, the right-hand side must be formed from Lorentz 
covariant tensors. In absence of any further structure, we have the 
metric $g_{\mu\nu}$ and the Levi-Civita symbol $\epsilon_{\mu\nu\kappa\lambda}$ 
at our disposal. The latter is excluded by the symmetry requirement 
$G_{\mu\nu}=G_{\nu\mu}$ following from the kinetic action. This leaves us with 
the ansatz,
\begin{equation}
 \{G_{\mu\nu}, G_{\kappa\lambda}\} = \big(a\, 
g_{\mu\nu}g_{\kappa\lambda}+b\, (g_{\mu\kappa} g_{\nu\lambda} + 
g_{\mu\lambda}g_{\nu\kappa})\big) \mathds{1},
\label{eq:GalgebraSM1}
\end{equation}
with constants $a,b$ to be determined. The metric factors have been arranged 
such that the symmetries of the anti-commutator and of $G_{\mu\nu}$ are already 
implemented. The identity $\mathds{1}$ in spinor space refers to the 
$d_\gamma\times d_\gamma$ matrix structure of the $G_{\mu\nu}$. 

The constants $a,b$ are determined by the requirements that the Luttinger 
operator should square to the square of the D'Alembertian,
\begin{equation}
 G_{\mu\nu} (i\partial^\mu) (i \partial^\nu) G_{\kappa\lambda} 
(i\partial^\kappa) (i \partial^\lambda) = (\partial^2)^2\quad \Rightarrow\quad 
a+2b=2,
\label{eq:Galgcond1}
\end{equation}
and that the $G_{\mu\nu}$ shall be traceless in order to remove the reducible 
Klein-Gordon part from the kinetic action,
\begin{equation}
 0=\{G_{\mu\nu},G^{\kappa}{}_\kappa\}=g^{\kappa\lambda}\{G_{\mu\nu}, 
G_{\kappa\lambda}\} \quad \Rightarrow\quad da+2b=0,
\label{eq:Galgcond2}
\end{equation}
where $d$ is the spacetime dimension. Solving these equations leads to 
\Eqref{eq:Galgebra}. By construction, this Abrikosov algebra is 
invariant under Lorentz transformations,
\begin{equation}
 G_{\mu\nu} \to G_{\kappa\lambda} \Lambda^\kappa{}_\mu \Lambda^\lambda{}_\nu,
\quad \Lambda\in \text{SO}(1,d-1) \label{eq:GLorentz}
\end{equation}
as well as under spin-base transformations \eqref{eq:spinbase} which correspond 
to the similarity transformations of the $G_{\mu\nu}$ matrices. (NB: in 
principle, we can perform GL($d_\gamma,\mathbb{C}$) transformations; however, 
the U(1) phase and a rescaling $\in \mathbb{R}_+$ does not change the 
$G_{\mu\nu}$ which is why we consider only SL($d_\gamma,\mathbb{C}$) 
\cite{Gies:2013noa,Gies:2015cka}.) For a given matrix $G_{\mu\nu}$, both 
transformations 
change the form of that matrix. However, since all representations are 
connected by similarity transformations \cite{Pauli:1936gd}, there exists a 
spin-base 
transformation $\mathcal{S}_{\text{Lor}}$ for each Lorentz transformation 
$\Lambda^{\mu}_{\nu}$, such that the spin-base transformation undoes the 
Lorentz transformation,
\begin{equation}
 G_{\mu\nu} \to \mathcal{S}_{\text{Lor}}G_{\kappa\lambda} \Lambda^\kappa{}_\mu 
\Lambda^\lambda{}_\nu \mathcal{S}_{\text{Lor}}^{-1} \equiv G_{\mu\nu}.
\label{eq:undo}
\end{equation}
Clearly, the set of all $\mathcal{S}_{\text{Lor}}$ forms an SO(1,$d-1$) subgroup
of SL($d_\gamma,\mathbb{C}$). For an accurate discussion of the global aspects
and depending on $d$, a restriction to the identity component may be necessary;
e.g. in $d=4$, the identity component of SO(1,3) has a universal cover that is
isomorphic to SL(2,$\mathbb{C}$), and therefore it is useful to think of
$\mathcal{S}_{\text{Lor}}$ as an SL(2,$\mathbb{C}$) subgroup embedded into
SL(32,$\mathbb{C}$), cf. below. The details of this embedding also quantify, how
the Luttinger spinor can be decomposed into SL(2,$\mathbb{C}$) Weyl spinors as
irreducible representations of the Lorentz group. The corresponding
transformations of the spinors $\psi \to \mathcal{S}_{\text{Lor}}\psi$ can be
viewed as the ``Lorentz transformations of Luttinger spinors''. This corresponds
to the conventional picture, where fields transform under Lorentz
transformations, but the $G_{\mu\nu}$ (or $\gamma_\mu$ in the Dirac case) remain
fixed. 

At this point, we should emphasize the difference between spin-base invariance 
and typical global symmetries such as the axial U(1) symmetry discussed for the 
NJL model in the main text. The latter is a typical ``Noether'' symmetry that 
acts on the fields and goes along with a corresponding Noether current and 
charge. By contrast, the spin-base symmetry also acts on the $G_{\mu\nu}$ which 
carries the spin structure but is not considered as a field in standard QFT. 
However, the $G_{\mu\nu}$ can be understood as fields $G_{\mu\nu}(x)$ in the 
case of gravitational interactions, where also the RHS of the 
Abrikosov algebra contains the metric $g_{\mu\nu}$ as the 
gravitational field variable. In such a scenario, the spin-base symmetry 
becomes a gauge theory rather than a global Noether symmetry 
\cite{Weldon:2000fr,Gies:2013noa}. 

For the metric convention $g=\text{diag}(1,-1,-1,\dots)$, the 
squares of the $G_{\mu\nu}$ can be determined from \Eqref{eq:Galgebra},
\begin{equation}
 G_{0i}^2=- \frac{1}{2} \frac{d}{d-1}, \quad G_{ij\neq i}^2= \frac{1}{2} 
\frac{d}{d-1}, \quad G_{\underline{\mu}\underline{\mu}}^2=1.
\end{equation}
Associating hermiticity properties to $G_{\mu\nu}$, this implies that $G_{0i}$ 
has to be chosen anti-hermitean and all others hermitean. Spanning the $G_{\mu\nu}$ 
matrices by a Euclidean Clifford algebra,
\begin{equation}
 G_{\mu\nu}= a_{\mu\nu}^A \gamma_A,\quad \{\gamma_A,\gamma_B\} = 2 \delta_{AB},
\end{equation}
with hermitean $d_\gamma\times d_\gamma$ matrices  $\gamma_A$, 
the coefficients $a_{0i}^A$ can be chosen imaginary, and all others real. With 
$G_{\mu\nu}=G_{\nu\mu}$ we need $d_e=\frac{1}{2}d(d+1)-1$ linearly independent 
anti-commuting elements to span the space of $G_{\mu\nu}$ matrices.

For illustration, let us give an explicit representation of the $G_{\mu\nu}$ 
matrices for $d=3+1$ dimensional spacetime in terms of $d_{\text{e}}=9$ Dirac 
matrices:
\begin{eqnarray}
 G_{0i}&=&i \sqrt{\frac{2}{3}} \gamma_{A=i}, \quad i=1,2,3, \nonumber\\
 G_{12}&=& \sqrt{\frac{2}{3}} \gamma_4, \quad G_{23} = \sqrt{\frac{2}{3}} 
\gamma_5, \quad G_{31}= \sqrt{\frac{2}{3}} \gamma_6, \nonumber\\
G_{00}&=& \gamma_7, \quad G_{11} = \frac{1}{3} \gamma_7 + \frac{2\sqrt{2}}{3} 
\gamma_8,\label{eq:Grep}\\
G_{22}&=& \frac{1}{3} \gamma_7 - \frac{\sqrt{2}}{3} \gamma_8 + 
\sqrt{\frac{2}{3}} \gamma_9, \nonumber\\
G_{33}&=& \frac{1}{3} \gamma_7 - \frac{\sqrt{2}}{3} \gamma_8 - 
\sqrt{\frac{2}{3}} \gamma_9. \nonumber
\end{eqnarray}
This representation can be viewed as an appropriate Wick rotation of   
the one constructed for $d=4$ Euclidean dimensions in \cite{Janssen:2015xga}. 

So far, it seems that the relativistic Abrikosov algebra could be 
constructed from the irreducible representation of the Euclidean Dirac algebra 
containing $d_{\text{e}}=9$ elements which would be the 
$d_{\gamma,\text{irr}}=2^{\lfloor \frac{d_{\text{e}}}{2}\rfloor}=16$ 
dimensional representation. However, a real action with a unitary time 
evolution requires the definition of the conjugate spinor $\bar\psi = 
\psi^\dagger h$, involving a spin metric $h$. In the Euclidean, $h=\mathds{1}$ 
can be chosen, as all $G$ matrices are hermitean. This is not a solution for the 
relativistic case, since the requirement of the action to be real implies 
the conditions of \Eqref{eq:hcomms}, $\{h,G_{0i}\}=0$, $[h,G_{ij}]=0$, 
$[h,G_{\underline{\mu} \underline{\mu}}]=0$, where we have used that 
$h=h^\dagger$ following from $\bar\psi\psi$ being a real scalar. The 
nonrelativistic choice $h=\mathds{1}$ is obviously in contradiction with 
$\{h,G_{0i}\}=0$. 

Let us for the moment assume that we work in the irreducible 
representation $d_{\gamma,\text{irr}}=16$. Using the representation 
\Eqref{eq:Grep} as an example, the latter anti-commutator can only be fulfilled, 
if $h$ is a (linear superposition of a)  product of an odd number of the 
remaining matrices $\gamma_{4, \dots,9}$ as this exhausts all possible elements 
of the algebra. Then, the resulting $h$ commutes only with $G_{ij}$ and 
$G_{\underline{\mu} \underline{\mu}}$, if it contains the elements $\gamma_A$ 
that $G_{ij}$ and $G_{\underline{\mu} \underline{\mu}}$ are composed from. This 
implies that $h$ must be a product of all remaining $\gamma_A$, e.g., 
$\gamma_{4}\gamma_{5}\gamma_6\gamma_7\gamma_8\gamma_9$. However, this is a 
product of an even number of $\gamma_A$ matrices in contradiction with the 
assumption made before. This proves that there exists no spin metric in the 
$d_{\gamma,\text{irr}}=16$ dimensional representation of the $\gamma_A$ 
matrices and thus no relativistic theory. 

The solution as presented in the main text is to use a reducible representation 
with the smallest one being $d_\gamma=32$ for $d=4$. Note that the attribute 
``reducible'' refers to being able to satisfy the Abrikosov algebra. 
If we read the algebra together with the conditions for the spin metric of 
\Eqref{eq:hcomms}, then the $d_\gamma=32$ dimensional representation cannot be 
further reduced to a lower dimensional representation without violating one of 
the requirements. Still, the reducibility in the aforementioned sense has a 
consequence: as the $d_\gamma=32$ dimensional representation goes along 
with two further anti-commuting elements $\gamma_{10}$ and $\gamma_{11}$, there 
are two linearly independent choices for a spin metric,
\begin{equation}
 h= \gamma_1\gamma_2\gamma_3\gamma_{10},\quad \text{or} \,\,\, 
\tilde{h}=\gamma_1\gamma_2\gamma_3\gamma_{11},
\label{eq:htildeh}
\end{equation}
satisfying all requirements of \Eqref{eq:hcomms}. Any linear combination 
$h'=\alpha h + \beta \tilde{h}$ with $\alpha^2+\beta^2=1$ and $\alpha,\beta\in 
\mathbb{R}$ could equally well serve as spin metric. At this point, the most 
important aspect is that a spin metric exists, rendering the action 
relativistically invariant, nonzero and real.

The reducibility (in the above mentioned sense) of the representation goes 
along with another advantage. There exists another element that anti-commutes 
with all $G$ matrices as well as with the spin metric. For our choice for the 
spin metric $h$, this element corresponds to $\gamma_{10}$ (for the choice 
$\tilde{h}$, it would be $\gamma_{11}$). Using this element, we can define 
axial transformations,
\begin{equation}
 \psi\to e^{i\vartheta \gamma_{10}} \psi, \quad  \bar\psi\to \bar\psi 
e^{i\vartheta \gamma_{10}}, 
\end{equation}
which leave the kinetic term invariant. Incidentally, a mass term of the 
form
\begin{equation}
 S_{m}= \int d^4x\, m^2 \bar\psi \psi
 \label{eq:Sm}
\end{equation}
would break this symmetry. The situation is therefore rather similar to the 
Dirac case with respect to both the existence of an axial symmetry in addition 
to the standard vector symmetry of phase rotations for the kinetic term, as 
well as the breaking of the axial symmetry by a mass term. This suggests that 
a Luttinger fermion can be decomposed into ``chiral'' components using the 
projectors
\begin{equation}
 P_{\text{R/L}}=\frac{1}{2} \left( \mathds{1}\pm \gamma_{10}\right), \quad 
\psi_{\text{R/L}}= P_{\text{R/L}} \psi.
\label{eq:chiralcomp}
\end{equation}
The kinetic term then decomposes into separate kinetic terms for the chiral 
components, whereas the mass term couples these components. This is analogous 
to the decomposition of Dirac fermions into Weyl components. We leave an 
analogous exploration of the existence of Majorana-Luttinger fermions or 
representations in terms of real Clifford algebras  \cite{Floerchinger:2019oeo} 
for future study.  

For the construction of possible interaction terms for the Luttinger 
fermions, it is useful to classify all linearly independent fermion bilinears 
of the form $\bar\psi \Gamma \psi$ with $\Gamma$ being a Clifford algebra 
element. The fact that $\psi$ is a 32-component spinor (together with reality 
requirements for the action) suggests that there might be $32\times32=1024$ 
bilinears. They can explicitly be listed using the Euclidean Dirac algebra. For 
this, we first note that $\gamma_{11} = \prod_{A=1}^{10} \gamma_A$. The complete 
set of Clifford algebra elements can thus be written as
\begin{equation}
 \Gamma=\{\mathds{1},\gamma_1,\gamma_2,\dots, \gamma_{10}, \gamma_1\gamma_2, 
\dots, \gamma_9\gamma_{10}, \gamma_1\gamma_2\gamma_3, \dots, \gamma_{11}\},
\end{equation}
listed in the form of an increasing length of the $\gamma$ products. Counting 
this number of products, yields
\begin{equation}
 \text{number of}\,\,\Gamma=\sum_{k=0}^{10} \left(\begin{array}{c} 10 \\ k 
\end{array}\right) = 1024, 
\label{eq:numberofGam}                                                       
\end{equation}
in agreement with the preceding expectation. It remains an interesting task to 
determine how these bilinears can be conveniently grouped into Lorentz tensors.

\section{Self-interacting quantum Luttinger fields}
\label{sec:AppB}

We have performed the computation of the one-loop $\beta$ functions for the 
purely fermionic theories using the functional RG, since the computational 
tools for such systems are fairly developed \cite{Gies:2001nw,%
Braun:2011pp,Gehring:2015vja}, and generalize 
straightforwardly to nonperturbative approximation schemes 
\cite{Hofling:2002hj,Braun:2010tt,Mesterhazy:2012ei,Jakovac:2014lqa,
Janssen:2014gea,Vacca:2015nta,Classen:2015mar,Knorr:2016sfs} to be explored in 
the future. We employ the Wetterich equation \cite{Wetterich:1992yh} for a 
scale-dependent effective action $\Gamma_k$,
\begin{equation}
 \pat \Gamma_k= \frac{1}{2} \STr \big[\pat R_k 
(\Gamma_k^{(2)}+R_k)^{-1}\big],
 \label{eq:Wetterich}
\end{equation}
where $R_k$ denotes a regulator function (specified below) in the Euclidean 
implementing a decoupling of low-momentum modes in the IR; its derivative $\pat
R_k$ establishes a Wilsonian momentum-shell integration, see 
\cite{Berges:2000ew,Pawlowski:2005xe,Gies:2006wv,Braun:2011pp} for reviews for 
the present setting. The supertrace $\STr$ includes a minus sign for all Grassmann-valued field components such as the Luttinger fermions considered here. For our purpose, it is useful to split the Hessian of the 
action into a field-independent kinetic part and a field-dependent part 
carrying all interactions, $\Gamma^{(2)}+R_k=\mathcal{P}+ \mathcal{F}$ with 
$\mathcal{P}=\Gamma^{(2)}_{\text{kin}}$ and 
$\mathcal{F}=\Gamma^{(2)}_{\text{int}}$. As a one-loop exact ansatz, we use, 
for instance, for the Gross-Neveu-type model of \eqref{eq:GrossNeveu}
\begin{equation}
\Gamma_k=\int_x \left[-Z\bar\psi G_{\mu\nu}\partial^\mu \partial^\nu \psi + 
\frac{\bar{\lambda}_0}{2} (\bar\psi\psi)^2 + \frac{\bar{\lambda}_t}{2} 
(\bar\psi G_{\mu\nu}\psi)^2\right],
\label{eq:GrossNeveuGamma}
\end{equation}
where we have introduced a wave function renormalization $Z$ which -- together 
with the bare couplings -- is considered as scale dependent. The RG flows of 
these quantities can be extracted upon an expansion of the Wetterich equation 
in powers of $\mathcal{F}$. The linear term contains the flow of the wave 
function which we parametrize in terms of the anomalous dimension of the 
fermion fields,
\begin{equation}
 \eta = - \pat \ln Z.
\end{equation}
In fact to one-loop order, we find $\eta=0$ which is consistent with the 
perturbative argument, that the only contributing tadpole diagram does not give 
rise to a nontrivial momentum dependence. For generality, we still keep the 
$\eta$ dependence explicit, in order to illustrate how higher-loop resummation 
effects would affect our results. The one-loop flow of the couplings is 
contained in the terms of order $\sim \mathcal{F}^2$. Introducing the 
renormalized (dimensionless) couplings
\begin{equation}
 \lambda_{0,t}= \frac{k^{d-4}}{Z^2} \bar{\lambda}_{0,t},
\end{equation}
we obtain the RG flows
\begin{equation}
 \pat \lambda_{0,t}= (d-4+ 2\eta)\lambda_{0,t}- 4 v_d l_2^{(d)}(0,\eta)\, 
f_{0,t}(\Nf,d,\lambda_0,\lambda_t),
\label{eq:patlambda0t}
\end{equation}
where we have kept the spacetime dimension general and used the abbreviation
$v_d^{-1}=2^{d+1}\pi^{d/2} \Gamma(d/2)$, such that  $v_4 = 1/(32\pi^2)$. The 
threshold function $l_2^{(d)}$ parametrizes the regularized loop-momentum 
integral 
also containing the information about the specifics of the regulator. In 
addition to the scaling terms linear in the couplings, the fluctuation-induced 
terms are defined in terms of the functions
\begin{eqnarray}
 f_0&=&(\Nf d_\gamma -2) \lambda_0^2  - \frac{2d}{d-1} (d_e+2) 
\lambda_0\lambda_t, \nonumber\\
f_t&=&- \frac{d}{(d-1)d_e} ((d_e-2)\Nf d_\gamma + 2 (d_e^2 -d_e +2)) \lambda_t^2 \nonumber\\
&&+\left(2 - \frac{4}{d_e}\right) \lambda_0\lambda_t - \frac{4}{d(d+2)} 
\lambda_0^2,
\label{eq:fg}
\end{eqnarray}
where we have used the abbreviation $d_e= \frac{1}{2} (d+2)(d-1)$. Note that 
both channels decouple in the large-$\Nf$ limit.

In general, the form of the threshold function depends on the chosen regulator 
$R_k$, manifesting the regularization scheme. For $d=4$, we can prove that the 
one-loop result is universal and independent of the scheme, as it should be. In 
order to arrive at an explicit expression, we choose a regulator that preserves 
spin-base invariance, $R_k(p)= G_{\mu\nu} p^\mu p^\nu\, r(p^2/k^2)$, where $r$ 
denotes a regulator shape function specifying the decoupling of IR modes. In 
fact, in the loop integrals, the shape function occurs in the combination $[p^2 
( 1+ r)]$ as can be expected from powercounting. The resulting loop structure 
can therefore be mapped onto that of bosonic threshold functions. In fact, the 
threshold functions $l_n^{d}(\omega,\eta)$ are well tabulated in the 
literature \cite{Braun:2011pp}, and the one in our equation above agrees with 
those. Using the common partially linear regulator shape function $r(y)= 
(y^{-1}-1) \theta (1-y)$ \cite{Litim:2001up}, we arrive at the standard result
\begin{equation}
 l_2^{(d)}(0,\eta)= \frac{4}{d} \left( 1- \frac{\eta}{d+2} \right),
 \label{eq:threshold}
\end{equation}
which in $d=4$ and at one-loop with $\eta=0$ boils down to 
$l_2^{(d=4)}(0,0)=1$. 
Specializing to $\Nf=1$, $d_\gamma=32$ and $d=4$, the flows of the couplings 
yield the expressions given in the main text. 

For the NJL-type model, the computation is performed analogously. Also in this 
case, the NJL coupling generates tensor-type couplings at subleading order in 
$\Nf$ which we ignore in our discussion. The flow equation for the NJL coupling 
can be brought into the form of \Eqref{eq:patlambda0t} with a corresponding 
function $f(\Nf,d, \lambda)=d_\gamma \Nf$ such that the flow equation for 
$d=4$ with $\eta=0$ reads
\begin{equation}
\pat \lambda = - \frac{d_\gamma \Nf}{8\pi^2}\, \lambda^2 \label{eq:NJLflow}
\end{equation}
in agreement with the special case for $d_\gamma=32$ given in the 
main text.

\section{Gauge theories for Luttinger fermions}
\label{sec:AppC}

Let us recall the action \eqref{eq:SLQED} for quantum electrodynamics (QED) with 
Luttinger fermions 
\begin{equation}\label{eqn:QEDaction}
    S = \int d^4x \left[-\frac 14 F_{\mu\nu} F^{\mu\nu} - \bar{\psi} G_{\mu\nu} 
\D^\mu \D^\nu \psi - m^2 \bar{\psi} \psi\right]. 
\end{equation}
The one-loop $\beta$ function can be derived via the effective action,  
integrating out the fermions in an electromagnetic background field
\cite{Dittrich:1985yb}. Apart from a normalization given below, the one-loop
correction to the action is given by the fermion determinant arising from
the integration over Grassmann-valued fields obeying Fermi-Dirac statistics:
\begin{eqnarray}  
    \Gamma_{1\ell}[A] &=& -i \ln\det [-G_{\mu\nu} \D^\mu \D^\nu - 
m^2]\label{eqn:Gamma} \\
    &=& -\frac i2 \ln\det [- G_{\mu\nu} \D^\mu\D^\nu G_{\kappa\lambda} 
\D^\kappa \D^\lambda+ m^4], \nonumber
\end{eqnarray}
where we used the properties $\gamma^2_{10}=1$ and $\{ 
G_{\mu\nu},\gamma_{10}\}=0$, in order to arrive at the squared differential 
operator -- similar to $\gamma_5$ hermiticity in standard QED 
\cite{Dittrich:1985yb}.

We can decompose the product $G_{\mu\nu} \D^\mu\D^\nu G_{\kappa\lambda} 
\D^\kappa \D^\lambda$ into symmetric and anti-symmetric parts, 
\begin{eqnarray}\label{eqn:s+as}
    && G_{\mu\nu} \D^\mu\D^\nu G_{\kappa\lambda} \D^\kappa \D^\lambda \\
    &&=\left(\frac 12 \{G_{\mu\nu}, G_{\kappa\lambda}\} + \frac 12 [G_{\mu\nu}, 
G_{\kappa\lambda}] \right) \D^\mu\D^\nu \D^\kappa \D^\lambda. \nonumber
\end{eqnarray}
Let us study the two pieces separately, starting with the anti-symmetric part,
\begin{equation} \label{eqn:GGDDDD}
    [G_{\mu\nu}, G_{\kappa\lambda}] \D^\mu\D^\nu \D^\kappa \D^\lambda = \frac 12 [G_{\mu\nu}, G_{\kappa\lambda}] [\D^\mu\D^\nu, \D^\kappa \D^\lambda],
\end{equation}
since $G_{\mu\nu} = G_{\nu\mu}$. The commutator between 
covariant derivatives yields
\begin{eqnarray} 
    [\D^\mu\D^\nu, \D^\kappa \D^\lambda]
    &=& -i e (F^{\nu\lambda} \D^\mu \D^\kappa + F^{\nu\kappa} \D^\mu \D^\lambda 
\nonumber\\
    &&+ F^{\mu\lambda} \D^\kappa \D^\nu + F^{\mu\kappa} \D^\lambda \D^\nu ).
\end{eqnarray}
Here we have used the relation $[\D^\mu,\D^\nu] = -ie F^{\mu\nu}$ and 
confined ourselves to  a constant electromagnetic field, $F^{\mu\nu} = 
\text{const.}$, such that $\D^\kappa F^{\mu\nu} =F^{\mu\nu} \D^\kappa $.

Since the product $[G_{\mu\nu}, G_{\kappa\lambda}] F^{\nu\lambda}$ is symmetric 
under the exchange $\mu \leftrightarrow \kappa$, \Eqref{eqn:GGDDDD} 
becomes 
\begin{equation}  \label{eqn:asp}
    [G_{\mu\nu}, G_{\kappa\lambda}] \D^\mu\D^\nu \D^\kappa \D^\lambda  
    = -ie [G_{\mu\nu}, G_{\kappa\lambda}]  F^{\nu\lambda} \{\D^\mu , \D^\kappa 
\}.
\end{equation}
%
The symmetric part of \Eqref{eqn:s+as} can be rewritten using the 
Abrikosov algebra \eqref{eq:Galgebra}; with the same assumptions, we obtain
\begin{equation}
    \{G_{\mu\nu}, G_{\kappa\lambda}\}
    \D^\mu\D^\nu \D^\kappa \D^\lambda 
    = 2 (\D^2)^2 +  \frac{3de^2}{2(d-1)} 
F_{\kappa\lambda}F^{\kappa\lambda},
\label{eqn:sp}
\end{equation}
where we work in general spacetime dimensions $d$ for generality, but will 
later specialize to $d=4$. Inserting Eqs.~\eqref{eqn:asp} and 
\eqref{eqn:sp} 
into \Eqref{eqn:s+as}, the one-loop effective action \eqref{eqn:Gamma} becomes
\begin{eqnarray} 
    \Gamma_{1\ell}[A] &= &-\frac i2 \ln\det \biggl[- (\D^2)^2 - \frac{3 e^2 
d}{4(d-1)} F_{\kappa\lambda}F^{\kappa\lambda} \nonumber\\
    &&+ \frac{ie}{2} [G_{\mu\nu}, G_{\kappa\lambda}]  F^{\nu\lambda} \{\D^\mu , 
\D^\kappa \} + m^4 \biggr].
\label{eqn:Gamma1loop} \end{eqnarray}
In order to keep track of potential issues arising from the tachyonic mass
poles or their negative residues, we work in Minkowski space, paying careful
attention to contour rotations in the complex momentum plane. First, we use the
Schwinger proper time formula for the logarithm,
\begin{equation}
    \ln{\frac{M}{N}} = - \lim_{\delta\to 0} \int_0^{\infty+i\delta} \frac{dt}{t} \left( e^{i
Mt} - e^{iNt} \right),
\end{equation}
such that \Eqref{eqn:Gamma1loop} can be written as
\begin{widetext}
\begin{equation}
     \Gamma_{1\ell}[A] =
     \frac i2 \lim_{\delta\to 0} \int_0^{\infty+i\delta} \frac{dt}{t} \Tr \biggl\{ e^{it\left[- (\D^2)^2 - \frac{3 e^2 d}{4(d-1)} F_{\kappa\lambda}F^{\kappa\lambda} + \frac{ie}{2} [G_{\mu\nu}, G_{\kappa\lambda}]  F^{\nu\lambda} \{\D^\mu , \D^\kappa \} + m^4 \right]} 
     - e^{it\left[- (\partial^2)^2 + m^4 \right]} \biggr\}.
\end{equation}
\end{widetext}
Here we have used $\ln\det (M)= \Tr\ln (M)$ and subtracted the 
free-field case in order to fix the normalization mentioned above. Since we aim 
at computing the $\beta$ function, it suffices to compute $\Gamma_{1\ell}$ to 
order $F^2$. For this, we define
\begin{eqnarray} 
    A &\equiv& - it(\D^2)^2 ,\nonumber\\
    B &\equiv& - \frac{et}{2} [G_{\mu\nu}, G_{\kappa\lambda}]  F^{\nu\lambda} 
\{\D^\mu , \D^\kappa \},
\end{eqnarray}
and employ the Baker–Campbell–Hausdorff formula for the expansion. We find
\begin{equation} 
    e^{A+B} = e^A e^B e^{-\frac 12 [A,B]} 
    \stackrel{\Tr}{\to} e^A (1+\frac 12 B^2) + \mathcal{O}(F^3),
    \label{eq:BCH}
\end{equation}
since $[A,B]^2 \sim \mathcal{O}(F^4)$, $B[A,B] \sim \mathcal{O}(F^3)$, 
and the traces of the terms $\sim B$ and $[A,B]$ vanish. This reduces our 
expression for the one-loop effective action to
\begin{align} 
     &\Gamma_{1\ell}[A] = \frac i2 \lim_{\delta\to 0} \int_0^{\infty+i\delta} 
\frac{dt}{t} e^{it\left[- \frac{3 e^2 d}{4(d-1)} 
F_{\kappa\lambda}F^{\kappa\lambda} + m^4 \right]}\,\times\nonumber\\ 
     & \Tr \!\left\{\! e^{-it (\D^2)^2 }\!\! \left[ \mathbb{1} + \frac 12 
\left( \!
\frac{-et}{2} [G_{\mu\nu}, G_{\kappa\lambda}] F^{\nu\lambda} \{\D^\mu , 
\D^\kappa \} \right)^2 \right] \!\right\} \nonumber\\ 
     &- \frac i2 \lim_{\delta\to 0} \int_0^{\infty+i\delta} \frac{dt}{t} \Tr 
\biggl\{ e^{it \left[- (\partial^2)^2 + m^4 \right]} \biggr\} + 
\mathcal{O}(F^4). \label{eqn:effact}
\end{align}
The functional trace runs over coordinate/momentum space as well as spinor 
space, $\Tr=\Tr_x \Tr_G$. Only the term $\sim B^2$ in \Eqref{eq:BCH} is 
nontrivial in spinor space, yielding
\begin{widetext}
\begin{equation} \begin{split} \label{eqn:traceB2}
     \Tr_G B^2 &= \frac{e^2t^2}{4} F^{\nu\lambda} \{\D^\mu , \D^\kappa \} F^{\beta\delta} \{\D^\alpha , \D^\gamma \} \Tr_G \biggl\{  [G_{\mu\nu}, G_{\kappa\lambda}] [G_{\alpha\beta}, G_{\gamma\delta}] \biggr\} \\
     &= \frac{e^2t^2}{4} \frac{8 d_\gamma}{(d-1)^2} [2d(2-d) F^{\nu\lambda} F_{\nu\kappa} \D_\lambda\D^\kappa\D^2 - d^2 F^{\nu\lambda} F_{\nu\lambda} (\D^2)^2],
\end{split} \end{equation}
where we have again kept terms only up to order $\sim F^2$, used 
$F=\text{const}.$, and $d_\gamma$ denotes the dimension of the Abrikosov 
algebra.
Plugging these results into \Eqref{eqn:effact}, we get to order $F^2$: 
\begin{equation} \begin{split} 
     \Gamma_{1\ell}[A] =& \frac{id_\gamma}{2} \lim_{\delta\to 0} 
\int_0^{\infty+i\delta} \frac{dt}{t} e^{itm^4} \left[1- i \frac{3 e^2 t 
d}{4(d-1)} F_{\kappa\lambda}F^{\kappa\lambda} \right] \Tr_x \biggl\{ 
e^{-it(\D^2)^2 } \\ 
    &  \times \left( 1 + \frac{e^2t^2}{(d-1)^2} [2d(2-d) 
F^{\nu\lambda} F_{\nu\kappa} \partial_\lambda\partial^\kappa\partial^2 - d^2 
F^{\nu\lambda} F_{\nu\lambda} (\partial^2)^2] \right)\biggr\} + 
\mathcal{O}(F^4) \\
     &- \frac{id_\gamma}{2} \lim_{\delta\to 0} \int_0^{\infty+i\delta} 
\frac{dt}{t} \Tr_x \biggl\{ e^{it \left[- (\partial^2)^2 + m^4 \right]} 
\biggr\}.
\end{split} \end{equation}
\end{widetext}
In order to use heat-kernel methods, we rewrite the factor $e^{-it (\D^2)^2 
}$ in terms of a Fresnel integral. For this, we use the Gaussian integral:
\begin{equation}
    \sqrt{\frac{\alpha}{\pi}}\int_{-\infty}^{+\infty} d\mu \ 
e^{-\alpha\mu^2-2\alpha\beta\mu} = e^{\alpha\beta^2},
\end{equation}
and implicitly rotate the contour by identifying $\alpha \equiv -i$ and 
$\beta^2 \equiv (\D^2)^2 t$, resulting in 
\begin{equation}
    e^{-it (\D^2)^2 } = \sqrt{\frac{-i}{\pi}}\int_{-\infty}^{+\infty} d\mu \ 
e^{i\mu^2}e^{2i\sqrt{t}\D^2\mu}. 
\end{equation}
The details of the contour rotation imply the 
convention $\sqrt{-i}=e^{-i\frac{\pi}{4}}$.
We are then left with the following integrals:
\begin{widetext}
\begin{equation} \begin{split} \label{eqn:3integrals} 
     \Gamma_{1\ell}[A]  = 
& \frac{id_\gamma}{2} \lim_{\delta\to 0} 
\int_0^{\infty+i\delta} \frac{dt}{t} e^{itm^4} \sqrt{\frac{-i}{\pi}} 
 \int_{-\infty}^{+\infty} d\mu \ 
e^{i\mu^2}\Tr_x \biggl[e^{2i\sqrt{t}\D^2\mu}  - e^{2i\sqrt{t}\partial^2\mu} 
\biggr]  \\ 
&+ \frac{id_\gamma}{2} \lim_{\delta\to 0} \int_0^{\infty+i\delta} \frac{dt}{t} 
e^{itm^4}  \Tr_x \biggl[ e^{-it(\partial^2)^2 } 
 \left( 
\frac{e^2t^2}{(d-1)^2} [2d(2-d) F^{\nu\lambda} F_{\nu\kappa} 
\partial_\lambda\partial^\kappa\partial^2 - d^2 F^{\nu\lambda} F_{\nu\lambda} 
(\partial^2)^2] \right.\\
& \qquad\qquad\qquad \left. - it \frac{3 e^2 d}{4(d-1)} 
F_{\kappa\lambda}F^{\kappa\lambda}\right) \biggr] + \mathcal{O}(F^4). 
\end{split} \end{equation}
\end{widetext}
Here, we use the heat kernel of the scalar Laplacian in a constant magnetic 
background field $B$ \cite{Gies:2002af},
\begin{eqnarray}
 \Tr_x e^{i\lambda \D^2}&=& -\frac{i\Omega}{(4\pi)^2} \frac{1}{\lambda^2} 
\frac{\lambda eB}{\sin \lambda eB} \\
&=& -\frac{i\Omega}{(4\pi)^2} \frac{1}{\lambda^2} \left(1+ 
\frac{\lambda^2e^2B^2}{6} + \mathcal{O}(B^4)\right)\nonumber, 
\label{eq:heatkernel}
\end{eqnarray}
where $\Omega$ denotes the spacetime volume, and the factor of $i$ arises from 
rotating the trace in momentum space into the Euclidean domain, $dp_0=idp_4$. 
The free-field subtraction cancels precisely the constant term in 
\Eqref{eq:heatkernel}. The remaining operators to be traced are all diagonal in 
momentum space and can be done straightforwardly. For instance, in terms of the 
magnetic background, the results read in our conventions
\begin{eqnarray} 
     \Tr_x && \biggl( e^{-it(\partial^2)^2 }  F^{\nu\lambda} 
F_{\nu\kappa}  (\partial^2)^2\biggr) = i \Omega \ 2B^2 
\int\frac{d^dp}{(2\pi)^d} e^{-ip^4t} p^4, \nonumber\\
     \Tr_x &&\biggl( e^{-it(\partial^2)^2 } F^{\nu\lambda} F_{\nu\kappa} 
\partial_\lambda\partial^\kappa\partial^2  \biggr)\nonumber \\
     &&= i \Omega \int\frac{d^dp}{(2\pi)^d} e^{-ip^4t} p_\lambda p^\kappa p^2 
F^{\nu\lambda} F_{\nu\kappa} \nonumber \\ 
&&=  \frac{i\Omega}{d} \ 2B^2 
\int\frac{d^dp}{(2\pi)^d} e^{-ip^4t} p^4, 
\end{eqnarray}
where the momentum integrals are understood to run over Euclideanized momenta. 
Because of the $t$ contour having a small positive imaginary part, the 
remaining integral converges, yielding, e.g., $\int \frac{d^4 p}{(2\pi)^4} 
e^{-i p^4 t} p^4 = - \frac{1}{32 \pi^2 t^2}$ in $d=4$.

Collecting the contributions of all terms, we arrive at 
\begin{equation} \begin{split}
    \Gamma_{1\ell}[A] & {=} - \frac{19}{18\pi^2} 
\frac{d_\gamma}{32}\Omega e^2B^2 \lim_{\delta\to 0} 
\int_{\frac{i}{\Lambda^4}}^{\infty+i\delta} \frac{dt}{t} e^{itm^4} \\
    & = \frac{19}{18\pi^2} \frac{d_\gamma}{32} \Omega e^2B^2 \left[ \gamma + 
\ln\left( \frac{m^4}{\Lambda^4}\right) + 
\Obig\left(\frac{m^4}{\Lambda^4}\right)\right],
\end{split} \end{equation}
to leading order $\mathcal{O}(B^2)$ in the field strength. By combining this 
result with the bare Maxwell Lagrangian using $\Gamma_{1\ell}= \Omega 
\mathcal{L}_{1\ell}$ for a homogeneous field, we get for the 
relativistic Luttinger case $d_\gamma=32$
\begin{eqnarray} 
    \mathcal{L}_\text{eff} &=& \mathcal{L}_\text{M} + \mathcal{L}_{1\ell} 
\nonumber\\
    &=& -\frac 12 B^2 + \frac{19}{18\pi^2} e^2B^2 \left[ \gamma + \ln\left( 
\frac{m^4}{\Lambda^4}\right)\right].
\end{eqnarray}
Defining the wave function renormalization
\begin{equation}
    Z^{-1} = 1- \frac{19}{9\pi^2} e^2 \left[ \gamma + \ln\left( 
\frac{\mu^4}{\Lambda^4}\right)\right],
\end{equation}
where $\mu$ is the RG scale, we introduce the renormalized field 
and coupling
\begin{equation}
    B_\text{R}^2 = Z^{-1} B^2, \quad e_\text{R}^2 = Z e^2.
\end{equation}
The one-loop $\beta$ function is then given by
\begin{equation}
    \beta_{e^2} \equiv \mu \frac{\partial}{\partial\mu} e_\text{R}^2(\mu) = 
\frac{\partial}{\partial\mu} Z e^2 = \frac{4 \cdot 19}{9\pi^2} e_{\text{R}}^4.
\end{equation}
By keeping track of which contribution arises from the Laplacian and which from 
the endomorphisms or the spin-field coupling terms, we can decompose the factor 
19 into dia- and paramagnetic contributions as is given in \Eqref{eq:betaQED} 
in the main text.  

The generalization to the nonabelian case is evident. We define  
the QCD action with one relativistic Luttinger quark by
\begin{eqnarray}
    S = \int_x \biggl[&&-\frac 14 F^a_{\mu\nu} F^{a\mu\nu} 
\label{eq:LQCDaction}\\
    &&- \bar{\psi}^i 
G_{\mu\nu} (\D^\mu)^{ij}(\D^\nu)^{jk} \psi^k - m^2 \bar{\psi}^i \psi^i \biggr],
\nonumber
\end{eqnarray} 
where $i,j=1,\dots, N_\text{c}$ labels fundamental and $a=1,\dots, 
N_\text{c}^2-1$ adjoint color indices. A generalization to arbitrary flavor 
numbers $N_{\text{f}}$ is straightforward. The covariant derivative is now given 
by
$
\D^\mu_{ij} = \partial^\mu - ig \tau^a_{ij} A^{\mu,a},
$
with $\tau^a$ being the generators of $\SU(N_\text{c})$, $\Tr_\text{c}(\tau^a 
\tau^b)= \frac 12 \delta^{ab}$.

The computation of the quark contribution to the QCD $\beta$ function can be 
mapped to that of the QED case, by using a pseudo-abelian background field
$A_\mu^a = n^a \tilde{A}_\mu$, where $\tilde{A}_\mu$ is an abelian vector 
potential and $n^a$ is a constant unit vector in color space ($n^an^a=1$).
The covariant derivative then reduces to 
$
    \D^\mu_{ij} = \partial^\mu - ig (\tau^an^a)_{ij} \tilde{A}^\mu,
$
which implies that the background is covariantly constant, 
$
    [D^\kappa, F^{\nu\lambda}] =
0.
$
The computation of the quark determinant proceeds in complete analogy to the 
QED case supplemented by the trace over color space. To leading order $\sim 
F^2$, this trace reduces to 
$\tr_{\text{c}}((\tau^an^a)_{ij}(\tau^bn^b)_{jk}) = \frac 
12 \delta^{ab} n^a n^b = \frac 12$, leading to a modification of the quark 
contribution to the QCD $\beta$ function by a factor of $\frac 12$ compared 
with QED, 
\begin{equation}
    \beta_{g^2}\Big|_{\text{quark-loop}} \equiv \mu 
\frac{\partial}{\partial\mu} g_\text{R}^2(\mu) = \frac{2 \cdot 19}{9\pi^2} 
g_{\text{R}}^4 ,
\end{equation}
for a single quark flavor. A different flavor number is accounted for by a 
factor of $N_{\text{f}}$. Together with the gluon (and ghost) loops, the final 
result for the QCD $\beta$ function with relativistic Luttinger quarks is given 
in \Eqref{eq:betaLQCD} in the main text.

\bibliography{bibliography}

\newpage
\newpage
\clearpage

\end{document}